\documentclass[10pt,aps,prl,twocolumn,groupedaddress,showpacs,longbibliography]{revtex4-1}

\usepackage[utf8]{inputenc} % set input encoding (not needed with XeLaTeX)
\usepackage[pdftex]{graphicx} % ou le fichier .pdf correspondant.
\graphicspath{{./Fig/}}
\usepackage{bm}
\usepackage{xcolor}
\usepackage[english]{babel}
\usepackage[
  colorlinks=true,
  urlcolor=blue,
  linkcolor=blue,
  citecolor=blue
]{hyperref}
\usepackage{array}
\usepackage[]{lineno}

\newcommand{\ee}{\mathrm{e}}
\newcommand{\ii}{\mathrm{i}}

\newcommand{\SM}{Supplementary Information}

\begin{document}
\title{Anomalous momentum diffusion in a dissipative many-body system} 

\author{Rapha\"{e}l Bouganne$^{1}$, Manel Bosch Aguilera$^{1}$, Alexis Ghermaoui$^{1}$, J\'{e}r\^{o}me Beugnon$^{1}$, Fabrice Gerbier$^{1}$}
\email[e-mail: ]{fabrice.gerbier@lkb.ens.fr}
\affiliation{$^{1}$Laboratoire Kastler Brossel, Coll\`{e}ge de France, ENS-PSL Research University, Sorbonne Universit\'{e}, CNRS, 11 place Marcelin-Berthelot, 75005 Paris}
\date{\today}

\maketitle 

\textbf{Decoherence is ubiquitous in quantum physics, from the conceptual foundations\,\cite{Zurek2003a} to quantum information processing or quantum technologies, where it is a threat that must be countered. While decoherence has been extensively studied for simple, well-isolated systems such as single atoms or ions\,\cite{Haroche2006}, much less is known for many-body systems where inter-particle correlations and interactions can drastically alter the dissipative dynamics\,\cite{Cai2013,Henriet2019,Poletti2013a,Pichler2010a}. Here we report an experimental study of how spontaneous emission destroys the spatial coherence of a gas of strongly interacting bosons in an optical lattice. Instead of the standard momentum diffusion expected for independent atoms\,\cite{Holland1996a}, we observe an anomalous sub-diffusive expansion, associated with a universal slowing down $\propto 1/t^{1/2}$ of the decoherence dynamics. This algebraic decay reflects the emergence of slowly-relaxing many-body states\,\cite{Poletti2013a}, akin to sub-radiant states of many excited emitters\,\cite{Henriet2019}. These results, supported by theoretical predictions, provide an important benchmark in the understanding of open many-body systems.
}

Interference phenomena are a central feature of quantum mechanics. However, they are easily destroyed by uncontrolled couplings with the environment, \textsl{i.e.} decoherence. In weakly correlated systems, inter-particle interactions are typically expected to hasten decoherence. For instance, they are responsible for the collisional broadening of spectral lines in hot atomic vapors. For strongly interacting many-body systems, the theory of non-equilibrium dynamics in general and of decoherence in particular remains challenging, and experiments can provide valuable insight. Improving our understanding of such problems could also help developing novel experimental methods harnessing dissipation to engineer specific quantum states\,\cite{daley2014a}.

Ultracold atoms provide a natural experimental platform to investigate these questions. Coherence of ultracold quantum gases is usually easily accessible experimentally, and the sources of relaxation are often well identified and experimentally controllable. Along these lines, experiments with dissipative atomic quantum gases have so far mainly explored the role of atom losses, demonstrating variants of the Zeno effect\,\cite{syassen2008a,barontini2013a,zhu2014a,tomita2017a,sponselee2018a}, bi-stability of transport\,\cite{labouvie2016a} or loss cooling\,\cite{rauer2016a,schemmer2018a}. Spontaneous emission provides a different dissipation mechanism. An atom excited by a near-resonant laser undergoes repeated photon absorption-spontaneous emission cycles. The atomic momentum changes randomly after each spontaneous emission and undergoes a random walk in momentum space with a width asymptotically scaling as $\Delta p \propto \sqrt{t}$\,\cite{Holland1996a,Pichler2010a}. This momentum diffusion, well-known in the context of laser cooling\,\cite{Wineland1979a,Gordon1980a}, suppresses interferences between different parts of the system, as observed in the pioneering experiment of \cite{pfau1994a}. The destruction of spatial coherence was also observed indirectly through the inhibition of tunneling for a dilute normal gas in an optical lattice\,\cite{patil2015a}, where interactions do not play any role. In addition, the impact of spontaneous emission on many-body localization has been recently studied experimentally\,\cite{luschen2017a}.

In this work, we study a quantum gas of strongly interacting bosons confined in optical lattices and decoherence is induced by applying a controlled rate of spontaneous emission. We observe that the presence of strong interactions between atoms leads to a dramatic modification of the time evolution of the momentum distribution. Whereas a normal diffusive evolution is expected for non-interacting particles, we observe a sub-diffusive behavior. We relate this behavior to a  slowing down of the loss of spatial coherence which, instead of the expected (fast) exponential decay for independent particles, shows a (slow) algebraic decay. Anomalous diffusion processes are common in classical statistical mechanics, describing many random walk processes with correlated or constrained steps\,\cite{Bouchaud1990a}. In particular, systems with a distribution of lifetimes featuring a long tail usually exhibit sub-diffusion with slower dynamics than the standard, uncorrelated random walk. The dissipative Bose-Hubbard model shares this feature\,\cite{Poletti2012a,Poletti2013a}, where slowly-relaxing states emerge when strong interactions shift the dissipative processes out of resonance. The ensuing sub-diffusive relaxation dynamics discovered in \cite{Poletti2013a} is consistent with the algebraic decay of the spatial coherence $\propto 1/t^{1/2}$ that we observe experimentally.

\begin{figure}[ht!]
\centering
\includegraphics[width=6cm]{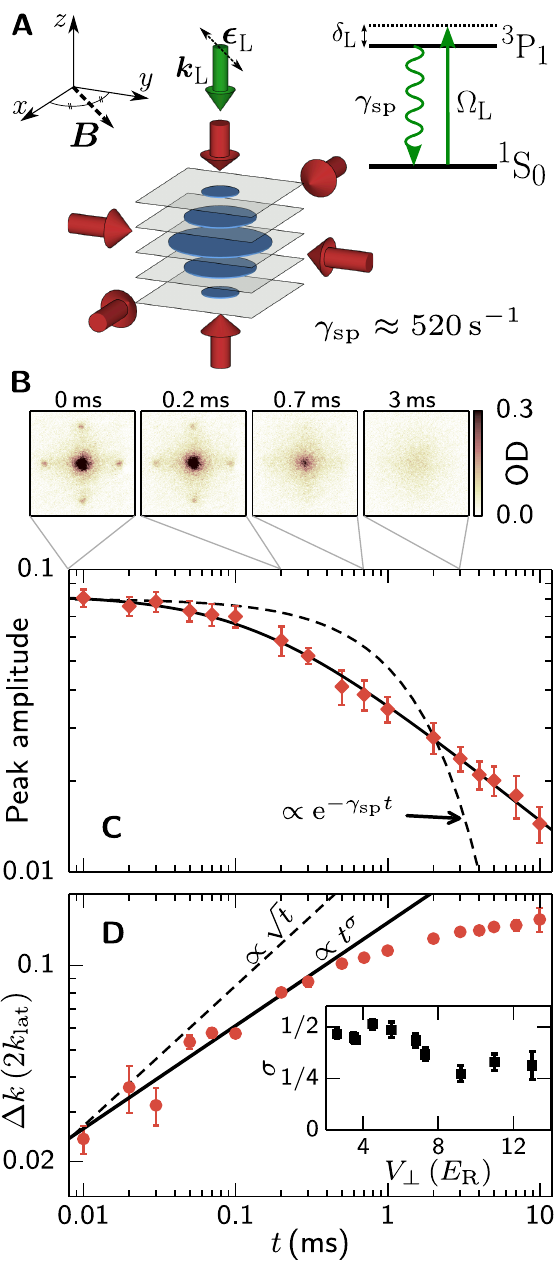}
\caption{\textbf{Observation of anomalous diffusion in momentum space.} (\textbf{\textsf{A}}) An ultracold gas of $^{174}$Yb atoms is trapped in a stack of two-dimensional square optical lattices and exposed to dissipation by spontaneous emission. A laser close to an atomic resonance (green arrow) induces fluorescence cycles at a rate $\gamma_{\mathrm{sp}} \approx 520\,$s$^{-1}$. Random recoil of the atom destroys the initial spatial coherence. (\textbf{\textsf{B}}) Absorption pictures showing the vertically integrated optical density (OD) after time of flight, revealing the momentum distribution $n({\bm k})$, for several dissipation times. (\textbf{\textsf{C}}) Time evolution of the peak amplitude of the momentum distribution. The dashed line shows the exponential decay expected for non-interacting atoms [Eq.\,(\ref{eq:expdecay})]. The solid line is a fit to $A/(1+\gamma_\mathrm{i} t/\kappa)^\kappa$. (\textbf{\textsf{D}}) Time evolution of the momentum width $\Delta k$. The observed saturation results from the finite size of the integration area. The short-time evolution follows a power-law (solid line) whose exponent $\sigma$ varies with lattice depth (inset). The dashed line indicates normal diffusion with $\sigma = 1/2$. In (\textsf{C}) and (\textsf{D}) the in-plane lattice depth is $V_\perp \approx 7.3\,E_\mathrm{R}$. Each point corresponds to the mean over 3 realizations of the experiment. Error bars are standard deviations of the mean.}
\label{fig1}
\end{figure}

In our experiments, we create degenerate quantum gases of bosonic $^{174}$Yb atoms trapped in a stack of independent, two-dimensional optical lattices (Methods and Fig.\,\ref{fig1}A). A (quasi-)condensate\,\cite{Bloch2008a} forms in each plane for small lattice depth $V_\perp$. When $V_\perp$ is roughly above $6\,E_\mathrm{R}$, the quantum gases are well described by a single-band Bose-Hubbard Hamiltonian\,\cite{Bloch2008a},
\begin{eqnarray}\label{eq:BH}
\hat{H}_{\mathrm{BH}} & = & - J \sum_{\langle i,j \rangle} \hat{a}_{i}^\dagger \hat{a}_{j} + \sum_{i}\left[ \frac{U}{2} \hat{n}_{i}(\hat{n}_{i}-1)  +V_i\,\hat{n}_{i} \right]. 
\end{eqnarray}
Here, $J$ is the tunneling energy between nearest neighbors, $U$ is the repulsive on-site interaction strength, $V_i$ is a harmonic potential arising from the Gaussian envelope of the lattice lasers\,\cite{Zwerger2003a,Bloch2008a}, and $\hat{a}_{i}$ and $\hat{n}_{i}=\hat{a}_{i}^\dagger \hat{a}_{i}$ are the annihilation and number operators for lattice site $i$. A superfluid-to-Mott-insulator transition occurs as $V_\perp$ and the ratio $U/J$ increase\,\cite{Bloch2008a}. The Mott insulator phase, where the atomic density is pinned at integer values, appears around $10\,E_\mathrm{R}$ for a filling of $\bar{n}=1$ atom per site. In this work, we explore a regime of lattice depths ranging from zero to $13\,E_\mathrm{R}$. The harmonic potential $V_i$ leads to an inhomogeneous spatial distribution with a maximum filling $\bar{n}\approx 2.5$ atoms per site.

We expose the atomic cloud to dissipation by shining a near-resonant laser beam for a given duration $t$ (Methods). Our main observable is the momentum distribution after a 20\,ms time-of-flight expansion. In the absence of dissipation, the momentum distribution corresponds to a multiple wave interference pattern, with sharp peaks at the Bragg positions where the matter waves interfere constructively\,\cite{Bloch2008a}. As shown in Fig.\,\ref{fig1}B, the Bragg peaks vanish rapidly (on a time-scale around $0.4 \,$ms $\sim 0.2\,\gamma_\mathrm{sp}^{-1}$) when dissipation is enabled. However, a residual structure in the momentum distribution persists for much longer times up to a few milliseconds. Besides the relaxation of coherence, we also observe atom losses that we attribute to two-body, light-assisted inelastic collisions\,\cite{weiner1999a}. We focus first on the evolution of coherence, and discuss the role of losses later below. In the remainder, we normalize the momentum distribution to the instantaneous atom number $N(t)$.

\begin{figure}[ht!]
\centering
\includegraphics[width=6cm]{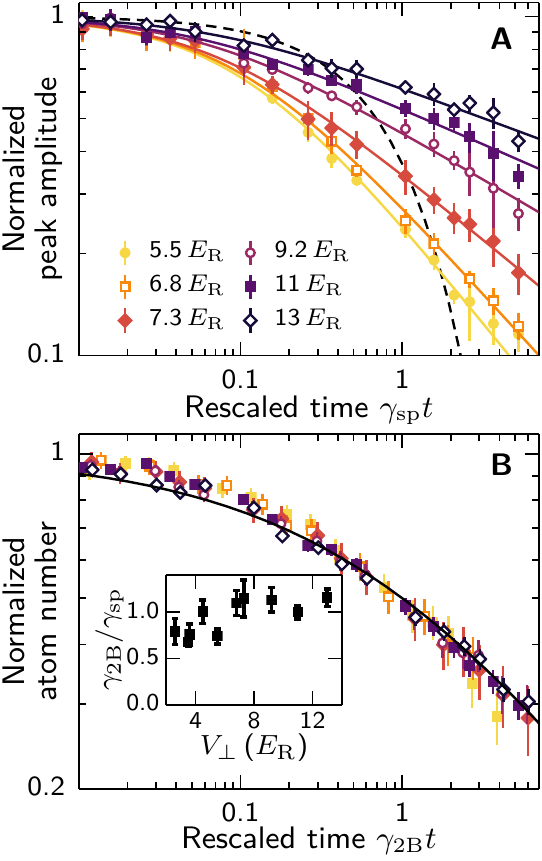}
\caption{\textbf{Decay of peak momentum amplitude and atom losses.} (\textbf{\textsf{A}}) Time evolution of the peak amplitude of the momentum distribution. The solid lines are fits as in Fig.\,\ref{fig1}. The dashed line shows the exponential decay expected for non-interacting atoms [Eq.\,(\ref{eq:expdecay})]. (\textbf{\textsf{B}}) Time evolution of the rescaled atom number $N/N_0$ versus rescaled time $x = \gamma_\mathrm{2B}t$. We extract the initial atom number $N_0$ and the two-body light-induced loss rate $\gamma_\mathrm{2B}$ (see inset) for each value of $V_\perp$ from a fit to $N_0/[1+x^\beta]$, with $\beta$ close to 1/2 for all lattice depths (\SM\ Fig.\,S3). The solid line is $1/(1+\sqrt{x})$. Each point corresponds to the mean over 3 realizations of the experiment. Error bars are standard deviations of the mean.}
\label{fig2}
\end{figure}

In order to characterize the decay of coherence, we plot in Fig.\,\ref{fig1}C the amplitude of the central Bragg peak versus time (Methods). The fast initial reduction is followed by a much slower power-law (``algebraic'') decay at long times. We show in Fig.\,\ref{fig2}A that this observation is valid for all lattice depths $V_\perp \geq 5\,E_\mathrm{R}$. For non-interacting atoms, one would expect that the quasi-momentum distribution $n^{(\mathrm{id})}$ relaxes exponentially to a uniform distribution equal to the mean number of atoms per site $\bar{n}$\,\cite{Pichler2010a,Yanay2014a} (\SM),
\begin{equation}\label{eq:expdecay}
n^{(\mathrm{id})}({\bm k},t) \approx n^{(\mathrm{id})}({\bm k},0) \mathrm{e}^{-\gamma_{\mathrm{sp}}t} + \bar{n}\left(1-\mathrm{e}^{-\gamma_{\mathrm{sp}}t}\right).
\end{equation}
Eq.\,(\ref{eq:expdecay}) predicts a faster decay at long times than experimentally observed, and cannot explain the power-law.

To reveal anomalous diffusion more directly, we compute the root-mean-square momentum width $\Delta k_t = \int_{\mathrm{BZ}_1} k_x^2 n({\bm k},t) \mathrm{d}^2 k$ from the images, where the integration is restricted to the first Brillouin zone $\mathrm{BZ}_1$. We show in Fig.\,\ref{fig1}D the momentum growth $\Delta k = \sqrt{\Delta k_t^2-\Delta k_{t=0}^2}$. The exponential decay in Eq.\,(\ref{eq:expdecay}) would lead to a normal law in $\Delta k \propto \sqrt{t}$ at short times, close to what we observe for small lattice depths (inset of Fig.\,\ref{fig1}D). However, in the Bose-Hubbard regime $V_\perp \geq 6\,E_\mathrm{R}$, we observe \textit{sub-diffusion} with a power-law behavior $\Delta k \propto t^\sigma$. The exponent $\sigma<1/2$ reaches $1/4$ at high lattice depths.

In the algebraic regime observed in Fig.\,\ref{fig2}A, the momentum distribution is characterized by a residual modulation on the scale of the first Brillouin zone, or equivalently by short-range spatial coherences before time of flight. Neglecting coherences beyond nearest neighbors, the lowest-band momentum distribution is approximately given by
\begin{equation}\label{eq:nk}
n({\bm k}) \approx \vert W({\bm k}) \vert^2 \Big( 1+ \sum_{{\bm d}=\pm {\bm e}_{x/y}} C_{\mathrm{nn}} \cos( {\bm k}\cdot {\bm d}) \Big).
\end{equation}
Here, the Wannier envelope $\vert W({\bm k}) \vert^2$ --the form factor-- reflects the on-site confinement, and the term between brackets --the structure factor-- is the residual interference pattern. The quantity $C_{\mathrm{nn}} = 1/N\sum_{\bm{r}_i} \langle \hat{a}_{\bm{r}_i+\bm{\delta}}^\dagger\hat{a}_{\bm{r}_i} \rangle$ is a spatially-averaged correlation function of the bosonic field between two nearest-neighbor sites at positions $\bm{r}_i$ and $\bm{r}_i+\bm{\delta}$, with $\bm{\delta}={\bm e}_{x/y}$ nearest-neighbor vectors of the square lattice.
 
\begin{figure*}[ht!]
\centering
\includegraphics[width=13cm]{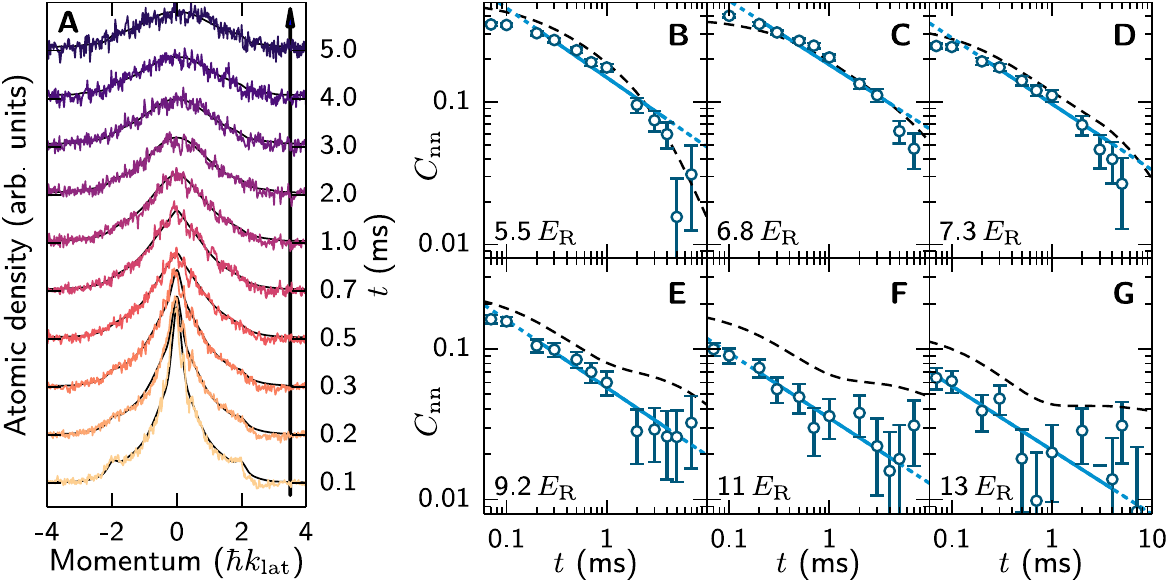}
\caption{\textbf{Decay of nearest-neighbour coherence.} (\textbf{\textsf{A}}) One-dimensional momentum profiles $n(\bm{k})$ for $\bm{k}=(k_x,0)$ versus dissipation time for a lattice depth $V_\perp \approx 7.3\,E_\mathrm{R}$. The dissipation time $t$ increases from bottom to top. Solid lines are fits to a multi-band function from which we obtain the nearest-neighbor coherence $C_{\mathrm{nn}}$. (\textbf{\textsf{B}} to \textbf{\textsf{G}}) Time evolution of $C_{\mathrm{nn}}$ for various lattice depths $V_\perp$. Solid lines are a fit to a power-law decay with exponent $\alpha$, extracted in a chosen time window (\SM). The dots show an extrapolation of the fit outside this window. The dashed lines show the prediction of the model described in the main text, including dissipation due to spontaneous emission and two-body light-induced losses. Error bars are 1-sigma confidence intervals derived from a $\chi^2$ fitting procedure.}
\label{fig3}
\end{figure*}

\begin{figure}[ht!]
\centering
\includegraphics[width=6cm]{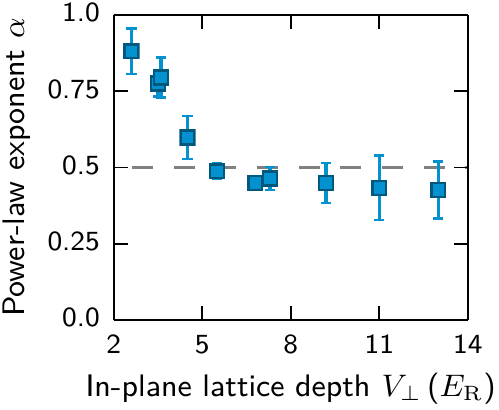}
\caption{\textbf{Decay exponent $\alpha$ of the nearest-neighbor coherence $C_{\mathrm{nn}}$.} Exponents are extracted from fits $C_{\mathrm{nn}} \propto 1/t^\alpha$, as shown in Fig.\,\ref{fig3}B-G, and error bars are 1-sigma confidence interval derived from statistical analysis of the fits. We observe a crossover from normal diffusion ($\alpha \approx 1$) to anomalous sub-diffusion with $\alpha \approx 1/2$.}
\label{fig4}
\end{figure}

We use a multi-band expansion analogous to Eq.\,(\ref{eq:nk}) to fit the measured momentum distributions and extract the fundamental band nearest-neighbor coherence $C_{\mathrm{nn}}$. We include the lowest excited bands to account for inter-band transitions induced by the excitation laser or spontaneous emission\,\cite{Pichler2010a} (Methods). Fig.\,\ref{fig3}A shows typical momentum profiles and the corresponding fits (also \SM\ Fig.\,S8). For lattice depths $V_\perp \geq 7\,E_\mathrm{R}$, we find that the nearest-neighbor coherence $C_\mathrm{nn}$ decays algebraically $\propto 1/t^{\alpha}$ (see Fig.\,\ref{fig3}B-G). For lower lattice depths a departure from the power-law is observed at long times. We plot the fitted exponent $\alpha$ in Fig.\,\ref{fig4} and find a crossover from $\alpha \approx 1$ for small $V_\perp$ to a plateau at $\alpha \approx 1/2$ for $V_\perp \geq 5\,E_\mathrm{R}$. This confirms the emergence of algebraic time relaxation and anomalous momentum diffusion in the Bose-Hubbard regime.

We now turn to the theoretical interpretation of our results. A first-principle description of the many-boson problem interacting with the quantized electromagnetic field is a difficult problem\,\cite{Pichler2010a}, and solving it without approximations a considerable task. Poletti \textit{et al.}\,\cite{Poletti2013a} have discussed a minimal single-band model where spontaneous light scattering is treated as a continuous, strictly local density measurement. For non-interacting atoms, the model reduces to the exponential relaxation described by Eq.\,(\ref{eq:expdecay}). Strong interactions drastically modify this relaxation process. The essence of the phenomenon can be traced back to the existence of states with low coherence but also low relaxation rates that dominate the long-times dynamics. 

The emergence of slowly-relaxing states is already seen in the simplest possible case with two atoms and two lattice sites. The ground state without dissipation is $\vert G \rangle \propto \vert 1, 1 \rangle - \sqrt{2}\epsilon\times(\vert 2,0 \rangle+\vert 0,2\rangle)$ to leading order in $\epsilon=J/U \ll 1$. Here $\vert n,m\rangle$ denotes a Fock state with $n$ atoms in the first well and $m$ in the second. When a weak dissipation $\gamma_{\mathrm{sp}} \ll U$ is enabled, the ground state $\vert G \rangle$ acquires a lifetime $\sim \gamma_\mathrm{sp} \epsilon^2\ll \gamma_\mathrm{sp}$: The interaction energy mismatch $\sim U$ between states with different occupation numbers shifts the dissipative processes out of resonance and strongly suppresses the relaxation. The other states are mostly superpositions of $\vert 2,0 \rangle$ and $\vert 0,2 \rangle$, for which dissipation randomizes the relative phase at the natural rate $\gamma_\mathrm{sp}$ (\SM).

Poletti \textit{et al.}\,\cite{Poletti2013a} have shown that such ``interaction-impeded decoherence'' persists with many atoms and many sites. Here a large number of slowly-relaxing states participate in the long-time dynamics when $\hbar \gamma_\mathrm{sp} \ll U$, leading to an algebraic regime with power-law decay for many observables. This dynamics is captured by a non-linear master equation governing the probability $p(n)$ to find $n$ atoms at a particular site. The transition rates become highly suppressed for configurations with a large interaction energy offset, and the distribution $p(n)$ obeys a scaling form underlying the algebraic behavior for small dissipation $\gamma_{\mathrm{sp}} \to 0$. From this scaling form derived in \cite{Poletti2013a}, one can show that $C_{\mathrm{nn}}$ obeys a universal law (\SM),
\begin{equation}\label{eq:Cnn_theo}
C_{\mathrm{nn}} \approx \frac{\eta}{\sqrt{z \gamma_\mathrm{sp} t}},
\end{equation}
with $\eta \approx 0.478$ a numerical factor and $z=4$ the number of nearest neighbors of a square lattice. The time evolution $\propto 1/t^{1/2}$ in Eq.\,(\ref{eq:Cnn_theo}) agrees well with our observations in Fig.\,\ref{fig4}, even though the system is inhomogeneous due to the auxiliary trapping potential. A numerical calculation using Gutzwiller theory and the local density approximation also confirms the survival of the $C_{\mathrm{nn}}\propto 1/t^{1/2}$ behavior in a trapped gas for high enough lattice depths (\SM). 

We finally discuss the role of atom losses in more details. In Fig.\,\ref{fig2}B, we show the time evolution of the normalized atom number $N(t)/N(0)$ for different lattice depths. We find that the data collapse onto the same curve $1/(1+x^\beta)$ with a scaling variable $x=\gamma_{\rm 2B}t$ (inset of Fig.\,\ref{fig2}B) and an exponent $\beta$ close to 1/2. This asymptotic scaling reflects the behavior of the two-body correlation function $\langle \hat{n}(\hat{n}-1)\rangle/\bar{n}^2 \propto 1/t^\beta$ (\SM). At long times, we expect that inelastic losses generate a gas of hardcore bosons with $\langle \hat{n}(\hat{n}-1)\rangle=0$, consistent with our measurements and reminiscent of the inhibition of losses in a 1D gas of molecules\,\cite{syassen2008a}. 

We have extended phenomenologically the theory of \cite{Poletti2013a} by adding a two-body loss term to the master equation for $p(n)$ (\SM). The characteristic two-body loss rate $\gamma_\mathrm{2B}$ is a free parameter adjusted to best match the atom number decay (Methods). We find a fair agreement with our data for the correlation function $C_\mathrm{nn}$ (Fig.\,\ref{fig3}, black dashed lines) and for the atom number decay up to $\gamma_\mathrm{sp}^{-1}$. We conclude that the lossy dynamics preserves the anomalous slowing down, only suppressing it when the fraction of lost atoms becomes large\,\cite{Sciolla2015}. At long times, the model fails to capture the observed dynamics. This could be explained by additional effects neglected in the theory, such as inter-band transitions, the dynamical creation of correlations between different sites or collective effects in light-matter interaction. For lattice depths $V_\perp \geq 10\,E_\mathrm{R}$, the model also overestimates the initial coherence, which could be due to a finite temperature of the sample.

Our experimental results can be summarized as the emergence of an algebraic decay law in a strongly-correlated quantum gas. Such power-law decay has been reported in several theoretical works exploring the dephasing of XY-spin chains\,\cite{Cai2013} or the influence of dipole-dipole interactions on optical clocks performances\,\cite{Henriet2019}. In all instances the power-law dynamics can be related to a slowly decaying, ``sub-radiant'' subspace dominating the long-time dynamics. Finding a common framework to describe this non-equilibrium dynamics, reminiscent of the classification of equilibrium phases into universality classes, provides an interesting, and to our knowledge open, question for future work.

\section*{Methods}

\subsection*{Optical lattices}
Our experiments are performed with a degenerate bosonic $^{174}$Yb gas of $7\times10^4$ atoms in a cubic optical lattice with spacing $d= \lambda_\mathrm{lat}/2$. The vertical confinement along gravity $V_{z} \approx 27\,E_\mathrm{R}$ is much stronger than the horizontal one, essentially freezing motion along $z$ and realizing a stack of independent two-dimensional quantum gases (see Fig.\,\ref{fig1}A). Here, $\lambda_\mathrm{lat}=2\pi/k_\mathrm{lat}\approx 760\,$nm is the wavelength of the lattice lasers, $E_\mathrm{R}=h^2/(2M\lambda_\mathrm{lat}^2)\approx h\times 1980\,$Hz the recoil energy, and $M$ the atomic mass. We calibrate the lattice depths along each axis independently using Kapitza-Dirac diffraction\,\cite{heckerdenschlag2002a}. 

\subsection*{Resonant excitation}
The dissipation laser operates near the so-called intercombination transition ${}^1\mathrm{S}_0 - {}^3\mathrm{P}_1$, of frequency $\omega_0=2\pi/\lambda_0$ and wavelength $\lambda_0 \approx 556$\,nm. The dissipation laser has wavevector $\bm{k}_\mathrm{L}$ and frequency $\omega_\mathrm{L} = c k_\mathrm{L}$ ($c$ is the speed of light in vacuum), propagates vertically and is detuned by $\delta_\mathrm{L} =\omega_\mathrm{L}-\omega_0= + 15 \Gamma_0$ from resonance, with $\Gamma_0=2\pi \times 180\,$kHz the excited state linewidth. The laser polarization ${\bm \epsilon}_\mathrm{L}$ is parallel to the uniform bias magnetic field $\vert{\bm B}\vert \approx 1\,$G (see Fig.\,\ref{fig1}A). The saturation parameter is $s \approx \Omega_\mathrm{L}^2/(2\delta_\mathrm{L}^2) \approx 10^{-3}$, with $\Omega_\mathrm{L}$ the Rabi frequency. The rate of spontaneous emission for a single atom in free space is then well-approximated by $\gamma_\mathrm{sp} \approx s \Gamma_0 / 2$, we calibrated its value using Rabi oscillations (\SM\ Fig.\,S1).

\subsection*{Analysis of peak amplitude and momentum width}
We estimate the peak amplitude in Figs.\,\ref{fig1} and \ref{fig2} from $n_{\mathrm{peak}} \equiv N_\mathrm{peak}/N$, where the total atom number $N$ (respectively, population $N_\mathrm{peak}$ of the central peak) is evaluated by counting the signal in a $480$-$\mu$m-wide square region around the atomic cloud (resp., $25$-$\mu$m-wide square in the centre of the image). For each $V_\perp$, we perform a fit using the phenomenological function $n_{\mathrm{peak}} = A/(1+\gamma_\mathrm{i} t/\kappa)^{\kappa}$, with $A$ the initial amplitude, $\gamma_{\mathrm{i}}$ the initial decay rate and $\kappa$ a decay exponent characterizing the long-time dynamics (\SM\ Fig.\,S6). The fit function interpolates between a linear decrease at short times and an algebraic decay at long times. The crossover time between the two regimes $\sim \kappa/\gamma_\mathrm{i}$, typically $\sim 0.1-0.2\,\gamma_\mathrm{sp}^{-1}$, is related to the disappearance of long-ranged spatial coherence.

\subsection*{Extraction of coherence from the momentum profiles}
Assuming negligible interactions and a far-field regime\,\cite{Pedri2001a,Gerbier2008a}, the time-of-flight distribution of a quantum gas released from an optical lattice mirrors the momentum distribution
\begin{equation}\label{eq:nk_allbands}
n(\bm{k}) = \sum_{\mathrm{bands}\,\bm{\nu}} \mathcal{S}_{\bm{\nu}}(\bm{k})\mathcal{W}_{\bm{\nu}}(\bm{k}).
\end{equation}
Here, the envelope function $\mathcal{W}_{\bm{\nu}}(\bm{k})$ is related to the Fourier transform of the Wannier function for each energy band labeled by $\bm{\nu}$. The normalized structure factor $\mathcal{S}_{\bm{\nu}}$ for band $\bm{\nu}$ is related to the correlation function $\langle \hat{a}_{\bm{\nu},i}^\dagger \hat{a}_{\bm{\nu},j} \rangle$,
\begin{equation}
\mathcal{S}_{\bm{\nu}}(\bm{k}) = \frac{1}{N}\sum_{i,j} \ee^{\ii\bm{k}\cdot(\bm{r}_i-\bm{r}_j)} \langle \hat{a}_{\bm{\nu},i}^\dagger \hat{a}_{\bm{\nu},j} \rangle.
\end{equation}
We truncate this expansion to the lowest band and the first few excited bands to model our experimental signal (see \SM\ for a more detailed account). We verified that including more terms in the expansion leads to negligible corrections. For the fundamental band, we write the structure factor as the sum of a ``coherent'' component $\mathcal{S}_{0,\mathrm{BEC}}(\bm{k}) $ describing the condensate, and of an ``incoherent'' component $\mathcal{S}_{0}(\bm{k})$ with only short-ranged coherence modelled by Eq.\,(\ref{eq:nk}). For the excited bands, which can be gradually populated by light scattering, we neglect coherence and take $\mathcal{S}_{\bm{\nu} \neq 0}(\bm{k}) = 1$. 

\subsection*{Theoretical model}
The calculations shown in Fig.\,\ref{fig3}B-G are performed using the master equation of \cite{Poletti2013a} for the on-site distribution $p(n)$, with an additional two-body loss term as discussed in the text (see \SM\ for more details). We compute the initial condition $p(n,\bm{r}_i)$ at each lattice site $\bm{r}_i$ by calculating the ground state of the Bose-Hubbard Hamiltonian using Gutzwiller theory\,\cite{Rokhsar1991a,krauth1992a} and a local density approximation to account for the harmonic potential $V_i$\,\cite{Zwerger2003a,Bloch2008a}. We then evolve $p(n,\bm{r}_i)$ using the master equation and finally average over the cloud to obtain the black dashed curves shown in Fig.\,\ref{fig3}B-G (also in \SM\ Figs.\,S4 and S5).

\section*{Acknowledgments}
We acknowledge fruitful discussions with A. Georges, C. Kollath and J.-S. Bernier. We thank M. Brune, J. Dalibard, R. Lopes, S. Nascimb\`ene and D. Poletti for careful reading of the manuscript.
\textbf{Funding:} LKB is a member of the DIM SIRTEQ of R\'egion Ile-de-France.
\textbf{Author contributions:} RB, MBA and AG performed the measurements under the supervision of JB and FG. RB analysed the data. RB and FG performed analytical and numerical calculations. All authors participated to the interpretation and discussion of the experimental results and to the writing the manuscript.
\textbf{Competing interests:} The authors declare that they have no competing interests.
\textbf{Data and materials availability:} All data needed to evaluate the conclusions in the paper are present in the paper and/or the Supplementary Information.

\bibliography{DissipativeBoseHubbard}
\bibliographystyle{apsrev_nourl}

\end{document}